\documentclass[
 reprint,
 superscriptaddress,
 amsmath,amssymb,
 aps,
 prl,
]{revtex4-2}

\usepackage{graphicx}
\usepackage{dcolumn}
\usepackage{bm}
\usepackage{hyperref}
\usepackage[mathlines]{lineno}
\usepackage{xcolor}
\usepackage{amsmath}
\usepackage{mathtools}

\begin{document}

\preprint{APS/123-QED}

\title{Spectroscopic Demarcation of Emergent Photons and Spinons\\ in a Dipolar-Octupolar Quantum Spin Liquid}

\author{Bin Gao}
 \thanks{These authors contributed equally to this work.}
 \affiliation{Department of Physics and Astronomy, Rice University, Houston, USA}
\affiliation{Rice Laboratory for Emergent Magnetic Materials and Smalley-Curl Institute, Rice University, Houston, TX 77005, USA}

\author{Zhengbang Zhou}
 \thanks{These authors contributed equally to this work.}
 \affiliation{Department of Physics, University of Toronto, Toronto, Ontario M5S 1A7, Canada}

\author{Tingjun Zhang}
 \affiliation{Department of Physics and Astronomy, Rice University, Houston, USA}
\affiliation{Rice Laboratory for Emergent Magnetic Materials and Smalley-Curl Institute, Rice University, Houston, TX 77005, USA}
\author{Andrey Podlesnyak}
 \affiliation{Neutron Scattering Division, Oak Ridge National Laboratory, Oak Ridge, Tennessee 37831, USA}

\author{Sang-Wook Cheong}
 \affiliation{Keck Center for Quantum Magnetism and Department of Physics and Astronomy, Rutgers University, Piscataway, New Jersey 08854, USA
}

\author{Yong Baek Kim}
\email{yongbaek.kim@utoronto.ca}
 \affiliation{Department of Physics, University of Toronto, Toronto, Ontario M5S 1A7, Canada}

\author{Pengcheng Dai}
\email{pdai@rice.edu}
 \affiliation{Department of Physics and Astronomy, Rice University, Houston, USA}
 \affiliation{Rice Laboratory for Emergent Magnetic Materials and Smalley-Curl Institute, Rice University, Houston, TX 77005, USA}

\date{\today}

\begin{abstract}
The identification of fractionalized excitations in quantum spin liquids (QSLs) remains a central challenge in condensed matter physics. In dipolar-octupolar (DO) pyrochlores, such as $\text{Ce}_2\text{Zr}_2\text{O}_7$, the candidate $\pi$-flux quantum spin ice (QSI) state is predicted to host both gapless emergent photons and a continuum of spinons. However, resolving these modes at zero field is complicated by their spectral overlap and the presence of nonmagnetic scattering near zero energy. Here, we report neutron scattering experiments on $\text{Ce}_2\text{Zr}_2\text{O}_7$ under a magnetic field along the $[1,1,1]$ direction. In contrast to previous unpolarized studies at 
zero-field that relied on high-temperature subtraction, we use a same-temperature high-field subtraction protocol to isolate the photon mode. Leveraging the selective coupling of the magnetic field to the dipolar degrees of freedom, we demonstrate the spectroscopic demarcation of these excitations. We observe that weak fields ($\approx 0.15$ T) suppress the low-energy photon weight while leaving the high-energy spinon continuum robust, albeit hardened. Our results, supported by gauge mean-field theory and exact diagonalization calculations, provide strong evidence for the $\pi$-flux QSI state and introduce a powerful field-tuning protocol for investigating DO-QSLs.
\end{abstract}

\maketitle


\begin{figure*}[t]
\includegraphics[width=0.95\textwidth]{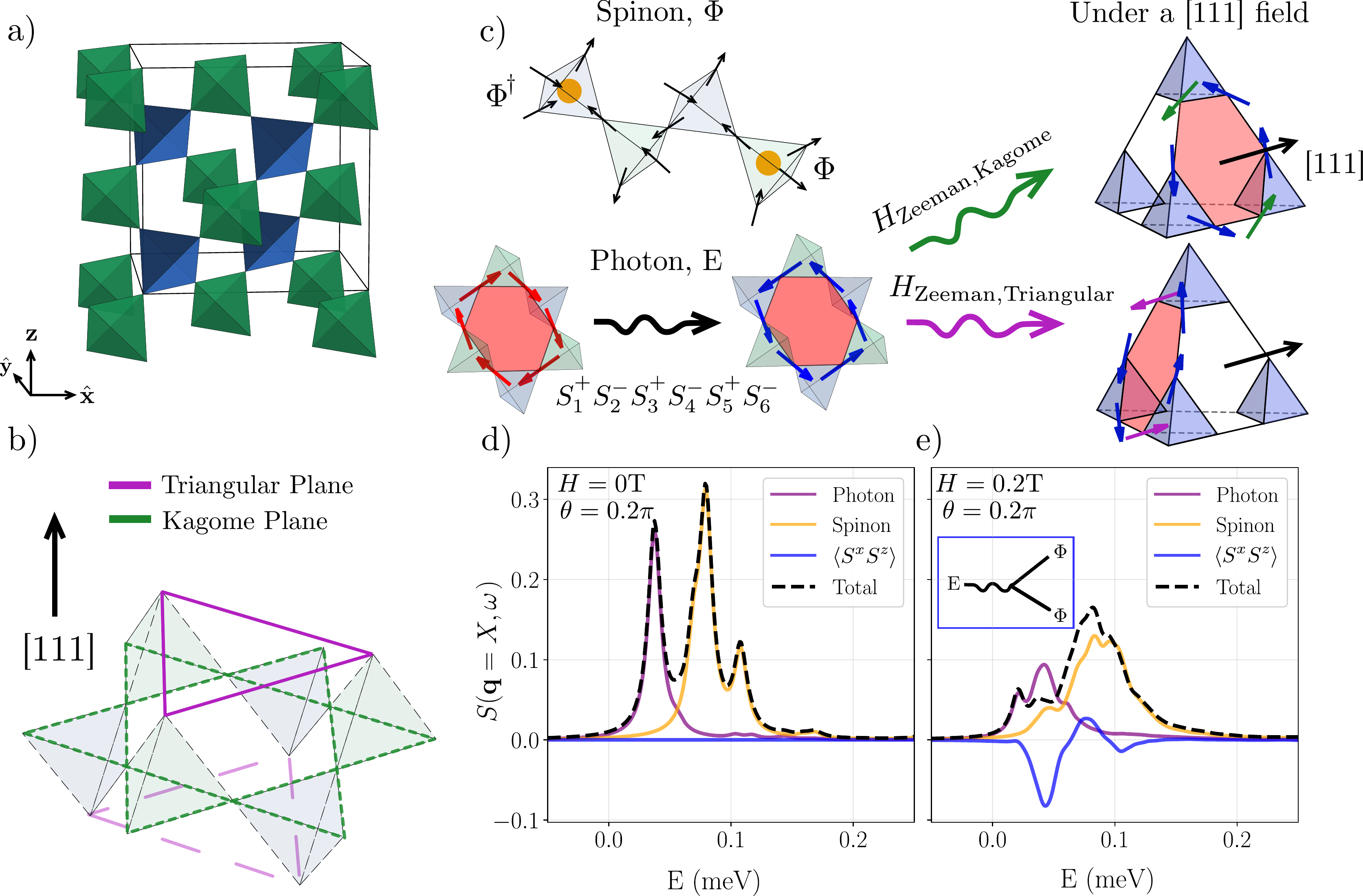} 
\caption{\label{fig:theory} (a) Pyrochlore network of Ce$^{3+}$ ions (corner-sharing tetrahedra) viewed along a cubic axis. (b) Decomposition of the pyrochlore lattice into Kagome planes (green) and triangular planes (magenta) stacked along the global $[1,1,1]$ direction (black arrow). In the DO basis, the Zeeman coupling in a $[1,1,1]$ field acts differently on these two types of layers, leading to layer-dependent gauge-field dynamics. (c) Spinon and photon sectors under a $[1,1,1]$ field. A gapped bosonic spinon $\Phi$ propagates on the dual diamond lattice formed by tetrahedron centers, with $\Phi^\dagger$ and $\Phi$ denoting creation and annihilation of gauge charges (orange circles) on opposite sublattices. The emergent photon $E$ corresponds to transverse fluctuations of the $U(1)$ gauge field on the six bonds of a pyrochlore hexagon; in the rotated local DO basis this is represented by the bond operator product $S_1^{+} S_2^{-} S_3^{+} S_4^{-} S_5^{+} S_6^{-}$, whose two circulation senses (red/blue arrows) give $\pm E$ excitations. Under a $[1,1,1]$ field, the Zeeman term couples to the dipolar component of the DO doublet and produces different photon dynamics on Kagome-layer bonds ($H_{\mathrm{Zeeman,Kagome}}$) and triangular-layer bonds ($H_{\mathrm{Zeeman,Triangular}}$). (d) ED calculation of the dynamical structure factor at the X-point, $S(\mathbf{q}=\mathrm{X},\omega)$, in zero field $H=0$ for a representative mixing angle $\theta=0.2\pi$. The total neutron response (black dashed) is decomposed into contributions from the emergent photon mode (purple), gapped spinon excitations (gold), and the photon-spinon mixing $\langle S^{x} S^{z} \rangle$ (blue). The dominant low-energy peak arises from the photon, while the spinon continuum lies at higher energy. (e) Same as (d) but in a finite $[1,1,1]$ field $H=0.2\ \mathrm{T}$. The Zeeman coupling distorts and partially gaps the photon branch, shifting and suppressing the low-energy photon peak, while the spinon contribution remains broad and comparatively robust. The inset illustrates the field-induced mixing between photon and spinon sectors in the GMFT spectrum: the eigenmodes acquire both $E$-like and $\Phi$-like character, enabling a field-tunable redistribution of spectral weight between photon and spinon channels that underlies our experimental disentanglement scheme.}
\end{figure*}

\begin{figure*}[t]
\includegraphics[width=0.95\textwidth]{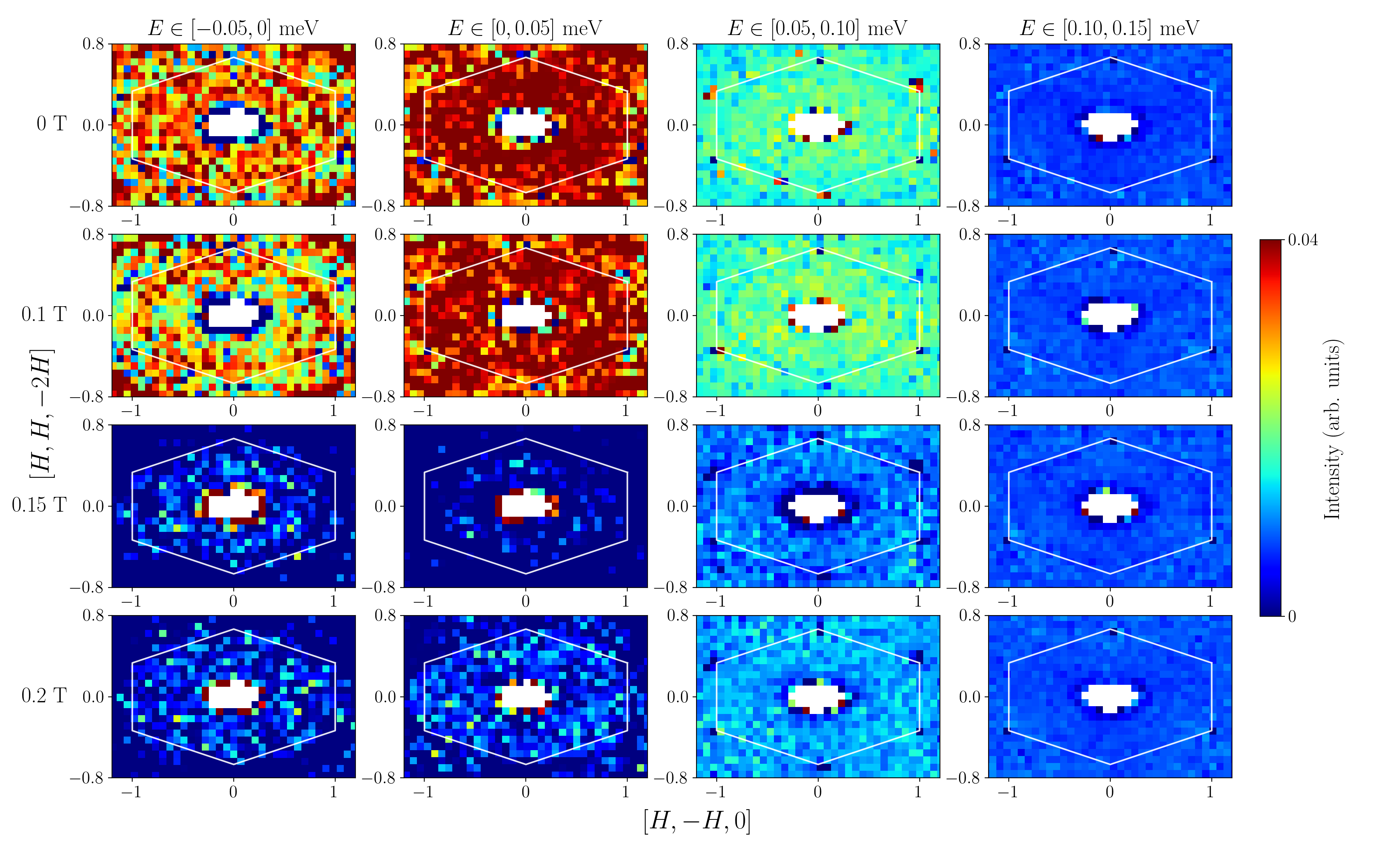} 
\caption{\label{fig:qmap} Field and energy evolution of the low-energy spectrum of Ce$_2$Zr$_2$O$_7$. Constant-energy slices of the inelastic structure factor $S(\mathbf{q},\omega)$ in the $([H,-H,0],[H,H,-2H])$ plane at $T \simeq 50$~mK, obtained by subtracting the same-temperature $3.0$~T data as a high-field background. Rows correspond to different $[1,1,1]$ fields, $\mu_{0}H = 0, 0.10, 0.15, 0.20$~T; columns show energy integration windows $E \in [-0.05,0]$, $[0,0.05]$, $[0.05,0.10]$, and $[0.10,0.15]$~meV, as indicated on the top. Hexagons mark the first Brillouin zone of the pyrochlore lattice in these reciprocal-lattice units. The first two columns are dominated by the emergent photon mode: intense, ring-like scattering at $H=0$ is rapidly suppressed by weak fields and is nearly absent by $0.15$~T. In contrast, the third column, which probes slightly higher energies, is dominated by the spinon continuum and remains relatively robust against the applied field. The rightmost column shows only weak, nearly field-independent residual intensity, indicating that the magnetic spectral weight is largely confined below $E \approx 0.10$~meV.}
\end{figure*}

\begin{figure}[t]
\includegraphics[width=0.9\columnwidth]{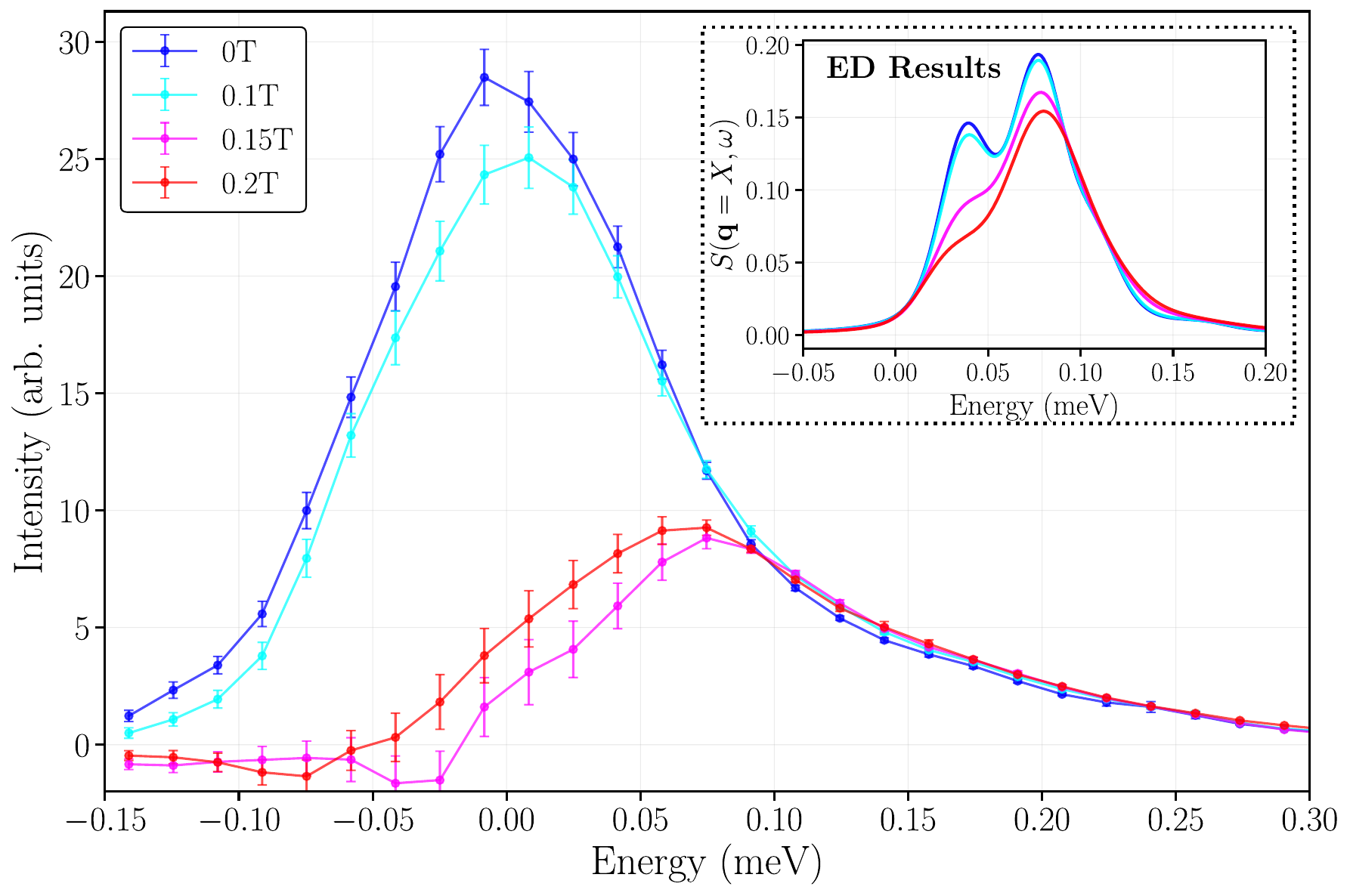} 
\caption{\label{fig:integrated} Field-induced redistribution of low-energy spectral weight. Main panel: Energy dependence of the momentum-integrated intensity in the first Brillouin zone of Ce$_2$Zr$_2$O$_7$ at $T \approx 50$~mK, after subtracting the same-temperature $3.0$~T dataset as a high-field background. Curves show data for $||\mu_{0}\mathbf{H}|| = 0, 0.10, 0.15, 0.20$~T (blue, cyan, magenta, red), integrated over the full $([H,-H,0],[H,H,-2H])$ plane excluding a region near the $\Gamma$ point. At zero field, a pronounced low-energy peak below $\sim 0.05$~meV reflects the emergent photon mode; with increasing field, this peak is rapidly suppressed, and its spectral weight is transferred toward the elastic line, while the higher-energy tail above $\sim 0.10$~meV remains comparatively unchanged. Inset: ED results for $S(\mathbf{q}=\mathrm{X},\omega)$ of the DO $U(1)$ QSL convolved with a Gaussian filter to emulate experimental resolution for the same set of fields, showing a similar evolution: the photon-dominated low-energy peak is weakened and broadened by the $[1,1,1]$ field, whereas the higher-energy spinon continuum is only weakly affected.}
\end{figure}

\textit{Introduction.}---Quantum spin liquids (QSLs) represent a frontier in condensed matter physics, offering a playground for long-range entanglement and fractionalized excitations \cite{Balents2010, Savary2017, wen2002quantum, wen2013topological, wen2004quantum, knolle2019field}. Within this landscape, the Quantum spin ice (QSI) state on the pyrochlore lattice is of particular interest. Here, quantum tunneling between degenerate ice configurations gives rise to an emergent compact $U(1)$ lattice gauge theory, predicted to host gapped bosonic spinons and, crucially, a gapless, linearly dispersing collective mode analogous to the photon of electromagnetism \cite{Hermele2004, Benton2012, tokiwa2018discovery}.

The Cerium-based dipolar-octupolar (DO) pyrochlores, $\text{Ce}_2{T}_2\text{O}_7$ ($T = \text{Sn, Zr, Hf}$), have recently emerged as prime candidates for realizing this $U(1)$ QSI state \cite{Gao2019, Sibille2015, Gaudet2019, Sibille2020, Smith2022, CZO227field, Poree2022, Beare2023, Poree2023, Smith2023}. These materials exhibit no long-range magnetic order or spin-glass freezing down to tens of millikelvin and display broad continua in inelastic neutron scattering (INS) consistent with fractionalization, all while maintaining minimal structural disorder \cite{Gao2019, Smith2022, CZO227field, Gaudet2019}. Theoretical analyses of the microscopic exchange parameters specifically place $\text{Ce}_2\text{Zr}_2\text{O}_7$ in the $\pi$-flux QSI regime \cite{Bhardwaj2022, Smith2022, Smith2023, smith2025twopeakheatcapacityaccounts}. Furthermore, recent comprehensive theoretical studies using gauge mean-field theory (GMFT) have mapped out the global phase diagram of these materials under magnetic fields \cite{Zhou2025}, predicting distinct signatures in the dynamical structure factor that can effectively disentangle the fractionalized excitations \cite{Zhou2024,bojesen2017Quantum}.

Despite these promising signs, definitively detecting the emergent photon---the ``light" of the QSI---remains a formidable experimental challenge. This mode resides in the quasi-elastic region near zero energy transfer ($E=\hbar\omega \approx 0$)~\cite{Benton2012, szabo2019seeing, Hermele2004}, where it is heavily obscured by the dominant nuclear incoherent and other nonmagnetic backgrounds. In previous unpolarized neutron scattering studies, a common strategy to remove this background was to subtract a high-temperature dataset (e.g., $T \approx 12$~K), under the assumption that the magnetic scattering at this temperature is diffusive and energy-independent; and nonmagnetic scattering is temperature independent between 30 mK and 12 K. However, as detailed in Ref.~\cite{Gao2025}, this method has some outstanding issues. The Bose population factor dictates that bosonic excitations within 1 meV—including phonons and other background scattering—are significantly populated at 12 K but suppressed at base temperature ($T \approx 30$~mK). Comparisons of the integrated intensity at the elastic line reveal that the 12 K signal is actually higher than the base-temperature signal across most of reciprocal space, likely due to thermally induced quasielastic scattering. Consequently, using the 12 K dataset as a background leads to a severe over-subtraction, causing the critical magnetic excitations near zero energy to be overlooked \cite{Gao2019, Sibille2015, Gaudet2019, Sibille2020, Smith2022, CZO227field}. Although neuron polarization analysis can resolve this problem \cite{Gao2025}, polarized neutrons have exceedingly low flux and cannot be carried out as a function of applied magnetic field, thus unable to determine the field dependence of the emergent photon and spinon scattering. 

In this Letter, we overcome this obstacle by employing a magnetic field applied along the $[1,1,1]$ direction as a precise spectroscopic tool. Leveraging the unique DO nature of the $\text{Ce}^{3+}$ pseudospins, where the field couples selectively to the dipolar ($\tau^z$) components while leaving the octupolar ($\tau^x, \tau^y$) sector largely unaffected \cite{Patri2020, huang2014quantum, Hosoi2022Uncovering,benton2020ground}, we implement a ``same-temperature high-field" background subtraction (STHFBS) protocol. Previous inelastic neutron scattering studies with fields along the $[1,\bar{1},0]$ and $[0,0,1]$ directions mainly tracked how the broad spinon-like continuum at zero field is suppressed as strong fields polarize the spins, without directly resolving the gapless photon mode \cite{CZO227field,Smith2023}, and these field geometries are not expected to cleanly disentangle photon and spinon spectral weight. Because the DO nature of Ce$_2$Zr$_2$O$_7$, we do not have field-induced ferromagnetic spin waves and thus can use the high-field state at the base temperature as an internal background reference. Following this STHFBS protocol, we show that a weak $[1,1,1]$ field efficiently suppresses the low-energy photon response while leaving the higher-energy spinon continuum largely intact. Combined with gauge mean-field theory (GMFT) and exact diagonalization (ED) calculations, we successfully extract the fragile emergent photon from the robust high-energy spinon continuum, providing independent and complementary spectroscopic evidence for the $\pi$-flux QSI state.

\textit{Experimental Methods.}---Inelastic neutron scattering experiments were performed using the Cold Neutron Chopper Spectrometer (CNCS) at the Spallation Neutron Source (SNS). A large single crystal of $\text{Ce}_2\text{Zr}_2\text{O}_7$ (mass $\approx 2$~g) was mounted on an oxygen-free copper holder. The sample was cooled to a base temperature of $T \approx 50$~mK using a dilution refrigerator equipped with a vertical-field cryomagnet, providing a magnetic field along the $[1,1,1]$ crystallographic direction. Data were collected at incident neutron energy $E_i = 3.32$~meV under five distinct magnetic fields: $\mu_0 H = 0, 0.1, 0.15, 0.2,$ and $3.0$~T. All experimental results presented in this work are shown after subtracting the $3.0$~T dataset as a background, effectively removing the field-independent nonmagnetic scattering. Full experimental details are provided in the Supplemental Material \cite{SI}.

\textit{Theoretical Calculations.}---%
%
We consider the most general nearest-neighbor model for DO pyrochlores~\cite{rau2019frustrated, huang2014quantum}, $
\mathcal{H} = \sum_{\langle i, j \rangle, \alpha} \tilde{J}_\alpha \tau^{\alpha}_i \tau^\alpha_j 
+ \sum_{\langle i, j \rangle} \tilde{J}_{xz} \left(\tau^x_i \tau^z_j + \tau^z_i \tau^x_j \right) 
- g_{zz}\mu_B\sum_i \left(\mathbf{H}\cdot \hat{\mathbf{z}}_i\right)\tau^z_i$,
with $\alpha\in\{x,y,z\}$. A rotation of the pseudospins about the local $y$ axis by an angle $\theta$, $\tau_i^y = S_i^y$,
$\tau_i^x = \cos\theta S_i^x - \sin\theta S_i^z$,
$\tau_i^z = \sin\theta S_i^x + \cos\theta S_i^z$,
diagonalizes the $(x,z)$ sector, with
$\tan(2\theta) = \frac{2\tilde{J}_{xz}}{\tilde{J}_x - \tilde{J}_z}$,
thereby eliminating the mixed $\tau_i^x \tau_j^z$ terms and yielding an XYZ Hamiltonian
\begin{align}
    \mathcal{H} &= \sum_{\langle i, j \rangle, \alpha} J_\alpha S^{\alpha}_i S^\alpha_j +\mathcal{H}_{\text{Zeeman}},
    \label{eq:H_XYZ}
\end{align}
where $\mathcal{H}_{\text{Zeeman}}=- \mu_B \sum_i g_{zz} (\mathbf{H} \cdot \hat{\mathbf{z}}_i) \left(\cos\theta S^z_i +\sin\theta S^x_i\right)$

In this XYZ basis, the low-energy physics is described by a compact $U(1)$ gauge theory coupled to bosonic spinons $\Phi$, obtained by identifying the spin component with the largest exchange as the emergent electric field~\cite{Hermele2004, Benton2012, szabo2019seeing}. For Ce$_2$Zr$_2$O$_7$, we adopt the experimentally motivated parameters
$(J_{xx}, J_{yy}, J_{zz}, \theta) = (0.063 \text{meV}, 0.062 \text{meV}, 0.011 \text{meV}, 0.2\pi)$,
for which $J_{xx}$ is dominant~\cite{Smith2022, Smith2023}. In the framework of GMFT~\cite{wen2002quantum,liu2019competing,liu2021symmetric,schneider2022projective,chern2021theoretical, chern2017fermionic,chern2017quantum, desrochers2023symmetry,savary2021quantum,desrochers2022competing,desrochers2023spectroscopic,Desrochers2024Finite, lee2012generic}, $S^x \sim E$ plays the role of the emergent electric field (photon), while the transverse components $
S^\pm = S^y \pm iS^z \sim \Phi^\dagger e^{iA}\Phi$
create and annihilate spinons in the presence of the gauge field $A$~\cite{desrochers2023symmetry, desrochers2022competing, savary2013spin, Balents2010, banerjee2008unusual, huang2018dynamics, pace2021emergent, Morampudi2020Spectroscopy}. Here we choose a larger $\theta$ for reasons discussed in the Supplemental Material \cite{SI}.

As previously discussed, neutron spins couple only to the dipolar component $\tau^z$, so the neutron–scattering intensity is proportional to the $\langle \tau^z \tau^z \rangle$ correlation function. In the $S^\alpha$ basis, $\langle \tau^z \tau^z\rangle 
= \sin^2\theta \langle S^x S^x\rangle 
+ \cos^2\theta \langle S^z S^z \rangle 
+ \sin(2\theta) \mathrm{Re}\langle S^x S^z\rangle$.
These three terms correspond, respectively, to the photon, spinon, and photon–spinon mixing channel. We evaluate the corresponding dynamical correlations using ED on a 16-site conventional cubic pyrochlore cluster with periodic boundary conditions~\cite{zhou2025quantumfisherinformationthermal, schafer2020pyrochlore, knyazev2001toward, gagliano1987dynamical, prelovsek2011ground}, which are compared with the experimental data.

\textit{Symmetry Consideration and Dynamics of Excitations}---Application of a magnetic field $\mathbf{H} \parallel [1,1,1]$ strongly modifies the dynamics of emergent photons. Because the local quantization axes $\hat{\mathbf{z}}_i$ differ on the four sublattices, the Zeeman term $\mathcal{H}_{\text{Zeeman}}$ couples to each sublattice with different strengths. A uniform $[1,1,1]$ field thus naturally decomposes the pyrochlore lattice into alternating Kagome and triangular layers, as shown in Fig.~\ref{fig:theory}(b). As established in previous work~\cite{sanders2024experimentally}, the resulting sublattice-dependent couplings split the photon dispersion into two branches, producing a broader and comparatively flatter spectral response, as illustrated in Fig.~\ref{fig:theory}(c). Our ED results confirm this behavior: the sharp photonic peak at zero field in Fig.~\ref{fig:theory}(d) continuously evolves into a broadened structure at $||\mu_0\mathbf{H}||=0.2$ T below the critical field in Fig.~\ref{fig:theory}(e).

A more subtle yet significant effect of the field is on the spinon–photon mixing channel. At zero field, the mixed correlator $\langle S^x S^z \rangle$ vanishes due to the residual $\mathbb{Z}_2$ symmetries corresponding to $\pi$ rotations generated by $S^x$ and $S^z$. In a finite $[1,1,1]$ field and for $\theta \neq 0$, the Zeeman coupling involves a linear combination of $S^x$ and $S^z$, explicitly breaking these symmetries and allowing $\langle S^x S^z \rangle$ to become finite. As shown in Fig.~\ref{fig:theory}(e), the resulting spinon–photon mixing term strongly suppresses the quasielastic spectral weight.

We stress that $\theta \neq 0$ is essential for two reasons. First, the photon contribution $\sin^2\theta\langle S^x S^x \rangle$ vanishes identically at $\theta = 0$, so a finite $\theta$ is required for the observed quasielastic response~\cite{Gao2025}. Second, within the quasielastic-suppression mechanism discussed above, if $\theta = 0$, the field couples purely to $S^z$, and a $\pi$ rotation generated by $S^z$ remains a good $\mathbb{Z}_2$ symmetry. In that case, $\langle S^x S^z \rangle$ is symmetry-forbidden, and the dramatic suppression cannot occur. The detection of the quasielastic peak at zero field and its subsequent suppression at finite fields are thus two mutually consistent manifestations of a finite mixing angle and the resulting spinon–photon mixing. In the following, we show that experimental data are consistent with these expectations.

\textit{Results and Spectral Demarcation.}---The evolution of the magnetic excitations is summarized in the momentum-space maps (${\bf q}$-maps) and energy cuts (Fig.~\ref{fig:qmap}). We categorize the response into three distinct energy sectors:

\textit{Photon sector $(E < 0.05\,\mathrm{meV})$.} At zero field, this window is dominated by a gapless, linearly dispersing mode~\cite{gingras2014quantum, Benton2012, Hermele2004}. Upon applying a weak $[1,1,1]$ field, the intensity is dramatically suppressed; by $||\mu_0\mathbf{H}|| = 0.15$ T, the signal in this range is nearly extinguished. While the data might be interpreted as a field-induced disappearance of photons, the larger critical field predicted by GMFT~\cite{Zhou2025} suggests otherwise. Rather, the combined effects of the $[1,1,1]$ field, discussed above, deplete the quasielastic signal and shift the photon-dominated spectral weight to finite energy as illustrated in Fig.~\ref{fig:theory}(d,e). In the neutron response, this redistribution mimics a gapped low-energy photon, even though the applied field remains below the true confinement transition.

\textit{Spinon sector $(0.05\,\mathrm{meV} < E < 0.10\,\mathrm{meV})$.} In this intermediate window, the intensity decreases with field but does not vanish. The reduction reflects the suppression of the high-energy tail of the photon mode, enhanced by the field-induced splitting of the photon branches. A robust residual signal, however, persists even at $||\mu_0\mathbf{H}|| = 0.2$ T. We associate this surviving component with the gapped spinon continuum, which remains dynamically active~\cite{Zhou2024, Zhou2025}.

\textit{The High-Energy Background $(E > 0.2\,\mathrm{meV})$:} This region shows negligible magnetic intensity across all fields, serving as an internal consistency check for our background subtraction method.

Crucially, we validate our experimental approach by comparing our zero-field integrated intensity results (Fig.~\ref{fig:integrated}) with the magnetic spectral weight isolated via polarized neutron scattering in Ref.~\cite{Gao2025}. The energy dependence obtained through our same-temperature high-field subtraction technique---characterized by a sharp, low-energy photon peak followed by a broader spinon continuum---is remarkably similar to the polarized energy scan which inherently separates magnetic from nonmagnetic scattering. This quantitative agreement confirms that both methods yield the correct intrinsic magnetic spectrum, establishing high-field subtraction as a robust and accessible alternative to complex polarized neutron setups for DO pyrochlores.

\textit{Discussion.}---The integrated intensity analysis (Fig.~\ref{fig:integrated}) reveals a redistribution rather than a simple loss of spectral weight. As the field increases, the integrated spectral weight over the first Brillouin zone, excluding a small region around the $\Gamma$ point, is strongly suppressed in the low-energy photon sector. Since the total magnetic spectral weight is conserved, we propose that this missing intensity is largely transferred to Bragg peaks (elastic scattering), consistent with the build-up of static correlations as the system polarizes.


The contrasting field evolution of the two components—a fragile quasielastic photon mode that is rapidly suppressed and a gapped spinon continuum that remains robust—constitutes a clear fingerprint of the composite nature of the QSI spectrum. In a conventional magnet, one expects that 
an applied field would induce a Zeeman gap 
in the spin wave spectra proportional to 
$||\mu_0\mathbf{H}||$ ($\approx 0.02$ meV at 0.2 T)
 without modifying the intensity of spin waves \cite{10.1093/oso/9780198862314.001.0001}. The dichotomous behavior observed here is difficult to reconcile with a simple magnon picture and instead naturally follows from the fractionalized gauge–spinon structure of the $\pi$-flux QSI state.

\textit{Conclusion.}---We have demonstrated that a $[1,1,1]$ magnetic field is a powerful tool for dissecting the excitation spectrum of the DO pyrochlore $\text{Ce}_2\text{Zr}_2\text{O}_7$. The field reshapes the quasielastic and inelastic response in a manner that allows us to delineate the fragile photon-dominated mode from the more robust spinon continuum through their distinct energy and field dependence. Our results not only provide compelling spectroscopic evidence for the $\pi$-flux QSI state but also validate the high-field background subtraction technique as a standard for future studies of QSL.

\begin{acknowledgments}
The single-crystal synthesis and neutron scattering experiments at Rice were supported by the U.S. DOE, BES under Grant Nos. DE-SC0012311 and DE-SC0026179 (P.D.). Part of the materials characterization efforts at Rice is supported by the Robert A. Welch Foundation Grant No. C-1839 (P.D.).  Z.Z. and Y.B.K. were supported by
the Natural Sciences and Engineering Research Council of
Canada (NSERC) Grant No. RGPIN-2023-03296 and the
Centre for Quantum Materials at the University of Toronto.
Computations at the University of Toronto were performed
on the Fir cluster, which is hosted by the Digital
Research Alliance of Canada. SWC was supported by the U.S. DOE, BES under Grant Nos. DE-FG02-07ER46382. This research used resources at the Spallation Neutron Source, DOE Office of Science User Facilities operated by the Oak Ridge National Laboratory (ORNL). ORNL is managed by UT-Battelle, LLC, under contract DE-AC05-00OR22725 with the U.S. DOE. The beam time was allocated to CNCS on proposal number IPTS- 34104.1.
\end{acknowledgments}


\clearpage
\appendix
\section*{Supplementary Information}
\setcounter{figure}{0}
\renewcommand{\thefigure}{S\arabic{figure}}
\setcounter{table}{0}
\renewcommand{\thetable}{S\arabic{table}}
\setcounter{equation}{0}
\renewcommand{\theequation}{S\arabic{equation}}

\section{Sample Preparation and Experimental Setup}
\begin{figure}
    \centering
    \includegraphics[width=0.8\linewidth]{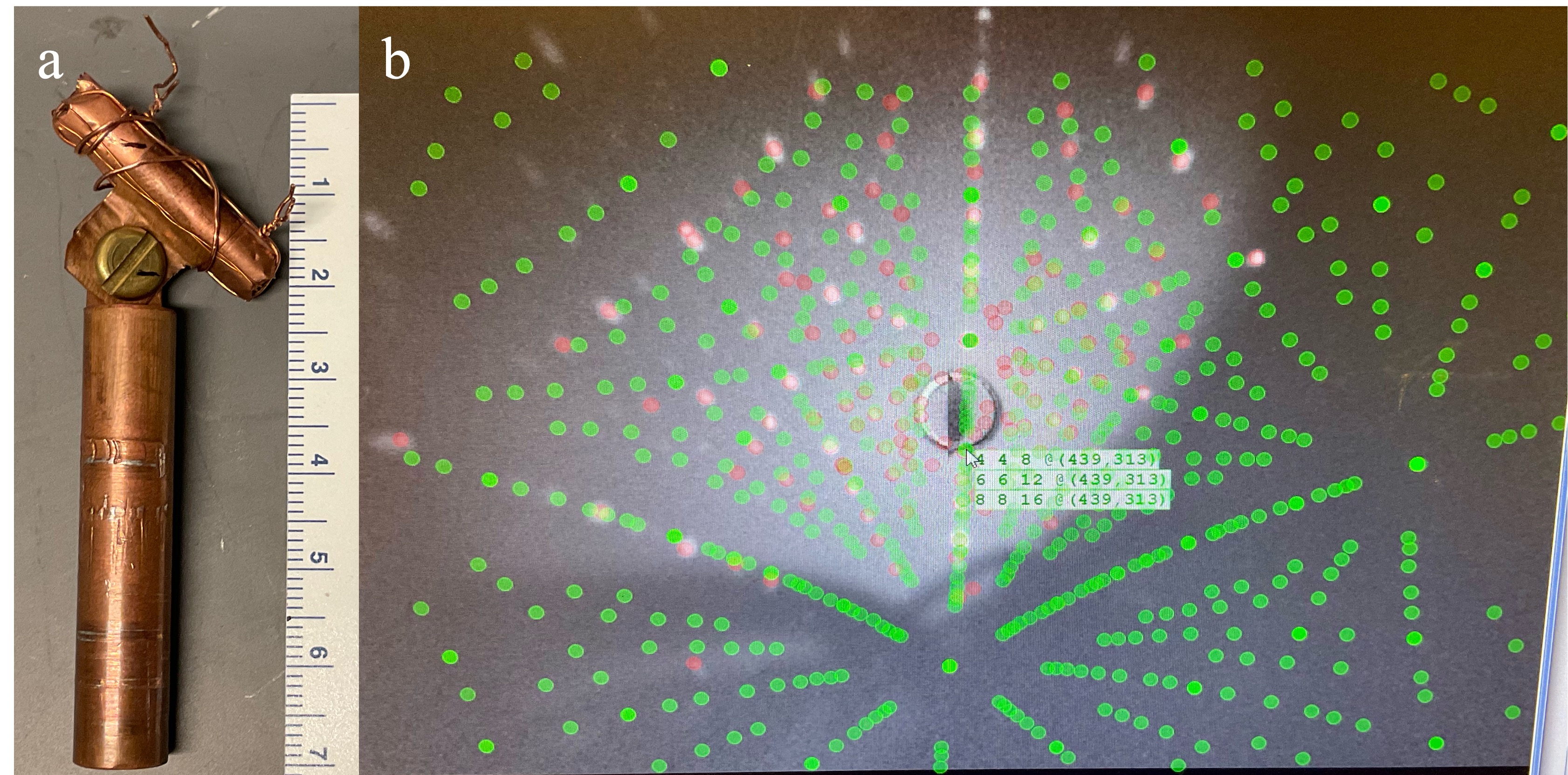}
\caption{\textbf{Sample preparation and alignment of Ce$_2$Zr$_2$O$_7$.}
\textbf{(a)} Photograph of the Ce$_2$Zr$_2$O$_7$ single-crystal specimen mounted for neutron-scattering measurements (copper holder/fixture shown with a ruler for scale). 
\textbf{(b)} Back-reflection Laue X-ray diffraction pattern collected from the same crystal, showing the orientation for subsequent scattering experiments, i.e.,  in the $[H, -H, 0] \times [K, K, -2K]$ scattering plane. White dots are the Laue spots, red dots are the spots recognized by the software, and green dots are the calculated spots from the software. The spot in the middle pointing by the mouse is [1, 1, 2] (or equivalent), with [1, 1, 1] in the vertical direction, and [1, 1, 0] in the horizontal direction. }
\label{fig:S_sample_laue}
\end{figure}
Single crystals of Ce$_2$Zr$_2$O$_7$ were grown using the optical floating zone technique, as described in detail in Ref.~\cite{Gao2019}. For the neutron scattering experiments, a single large crystal with a mass of approximately 2 grams was used. The orientation was verified using X-ray Laue diffraction (Fig. \ref{fig:S_sample_laue}).

The inelastic neutron scattering experiments were conducted at the Cold Neutron Chopper Spectrometer (CNCS) at the Spallation Neutron Source (SNS), Oak Ridge National Laboratory (ORNL). The sample was mounted on an oxygen-free copper holder and aligned in the $[H, -H, 0] \times [K, K, -2K]$ scattering plane. The sample environment consisted of a dilution refrigerator inserted into a vertical-field cryomagnet, enabling measurements at a base temperature of $T \approx 50$ mK and magnetic fields up to 8.0 T applied along the vertical $[1,1,1]$ direction.

We utilized an incident neutron energy of $E_i = 3.32$ meV, which provided an elastic energy resolution of approximately $0.1$ meV (full width at half maximum) at the elastic line. The sample was rotated over a wide angular range of $180^\circ$ in steps of $1^\circ$. Data were reduced using the \textsc{Mantid} software package and analyzed using \textsc{Dave}. We did not symmetrize the data to improve the counting statistics. Instead, we only applied a 180° rotation to replicate the measured angular wedge into the opposite half of the reciprocal space, providing full reciprocal space coverage without symmetry averaging.

\section{Raw data and the comparison to polarized neutron
\label{sec:rawdata}}
\begin{figure}
    \centering
    \includegraphics[width=0.8\linewidth]{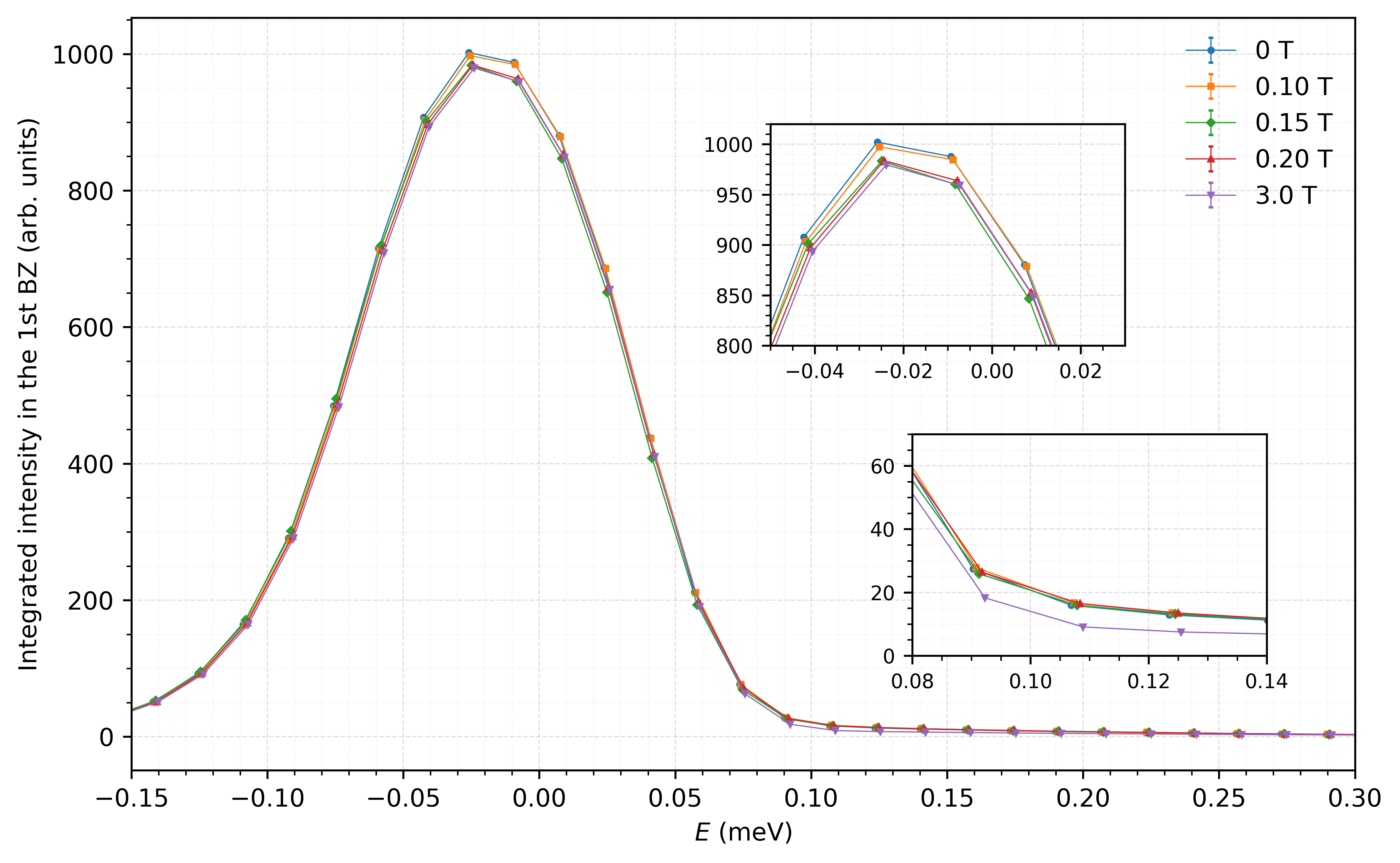}
    \caption{\textbf{Raw data of the integrated 1st-Brillouin-zone $E$-cut in Ce$_2$Zr$_2$O$_7$.}
Energy-dependent neutron-scattering intensity integrated over the first Brillouin zone in the
$[H, -H, 0] \times [K, K, -2K]$ scattering plane for five applied magnetic fields (0, 0.10, 0.15, 0.20, and 3.0~T).
Points show the raw integrated intensities with statistical error bars.
The upper inset zooms into the quasi-elastic peak region (photon window), highlighting the field-dependent changes near $E\approx 0$~meV.
The lower inset magnifies the gaped spinon window to emphasize the field dependence near $E\approx 0.1$~meV.}
    \label{fig:S_rawdata}
\end{figure}

To quantify the field dependence of the low-energy spectral weight in Ce$_2$Zr$_2$O$_7$, we constructed $E$-cuts by integrating the scattering intensity over the first Brillouin zone (BZ) in the $[H, -H, 0] \times [K, K, -2K]$ plane. The background-subtracted data obtained using our same-temperature high-field background subtraction (STHFBS) protocol (with the 3.0~T dataset as the reference) are plotted in Fig.~3 of the main text.

\textbf{STHFBS (3.0~T reference): raw vs.\ background-subtracted spectra.}
The 1st-BZ--integrated $E$-cuts shown in Fig.\ref{fig:S_rawdata} are \emph{raw} intensities and therefore contain a large, predominantly \emph{non-magnetic background} (nuclear/incoherent scattering, phonons, resolution tails, and instrumental contributions) in addition to any magnetic signal. To isolate the low-energy magnetic response, we apply the STHFBS protocol and define 
\begin{equation}
I_{\mathrm{mag}}(E)=I_{\mathrm{raw}}(E;H)-I_{\mathrm{raw}}(E;3~\mathrm{T}),
\end{equation}
which removes the largely field-independent background and reveals the small, field-dependent gapless spectral weight (photon window) and the gaped spectral weight (spinon window).
\begin{figure}
    \centering
    \includegraphics[width=0.8\linewidth]{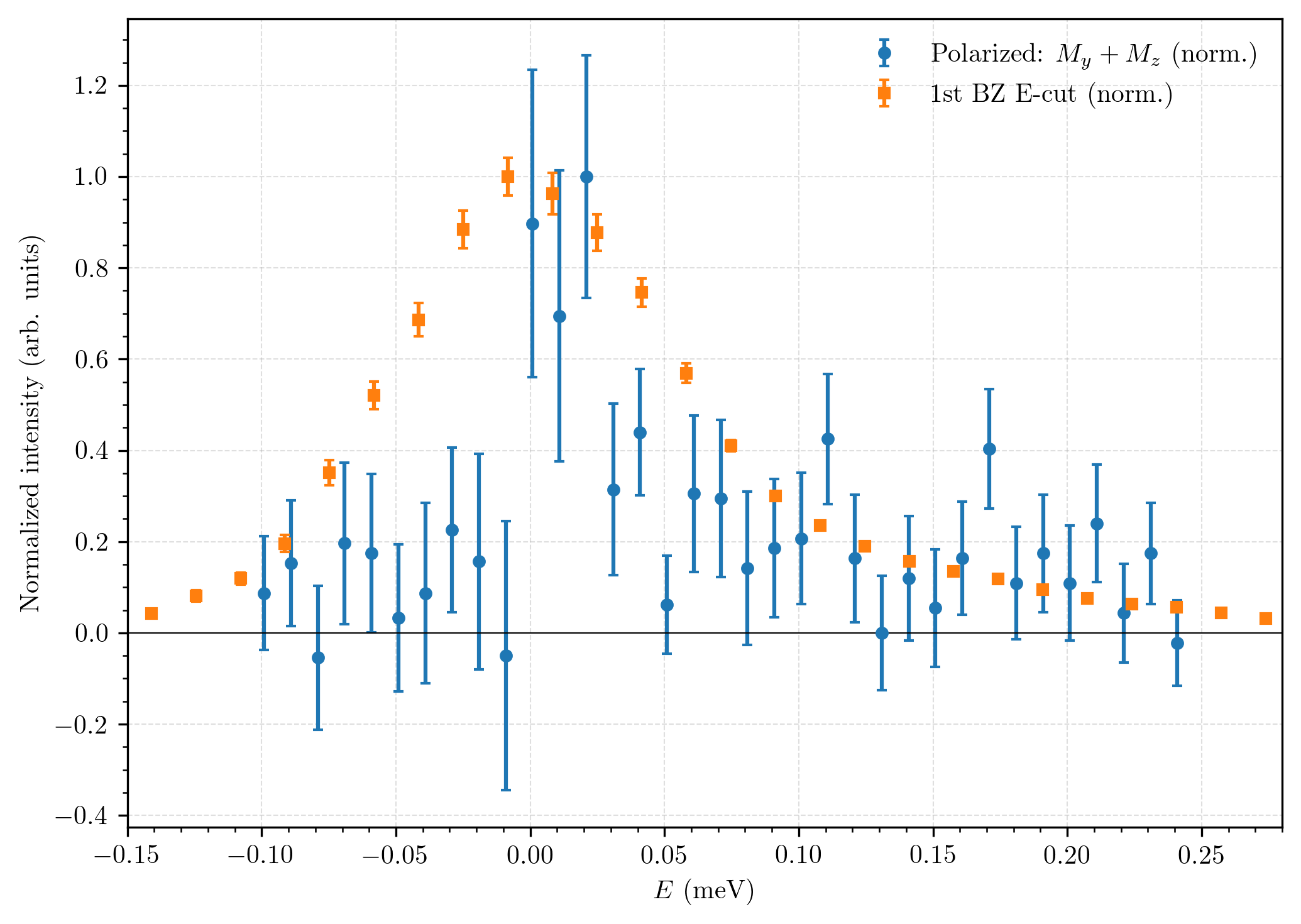}
    \caption{\textbf{Comparison between the background-subtracted 1st-BZ $E$-cut and the polarized-neutron $M_y\!+\!M_z$ spectrum.}
Orange squares: normalized $E$-cut obtained from the present field experiment using the STHFBS protocol, $I_{\mathrm{mag}}(E)=I_{\mathrm{raw}}(E;0~\mathrm{T})-I_{\mathrm{raw}}(E;3.0~\mathrm{T})$, after integrating the intensity over the first Brillouin zone in the $[H, -H, 0] \times [K, K, -2K]$ plane. Blue circles: normalized magnetic response $M_y(E)+M_z(E)$ extracted from polarized-neutron spin-flip channels (with standard polarization analysis) at $X = [0, 0, 1]$ point reported in Ref. \cite{Gao2025}. Both datasets are shown with statistical uncertainties and are normalized (each curve scaled to its own maximum) to facilitate comparison of the spectral shape.}
\label{fig:S_compare_pol_vs_field}
    \label{fig:placeholder}
\end{figure}

\noindent\textbf{Photon fraction relative to the raw signal.}

Quantitatively, the photon signal represents only a few percent of the raw integrated intensity near $E\approx 0$~meV. From the 1st-BZ integrated spectra, $I_{\mathrm{raw}}\sim 10^{3}$ (arb.\ units) while the subtracted low-energy magnetic contribution is $\sim 2\times 10^{1}$--$3\times 10^{1}$ (arb.\ units), giving
\begin{equation}
\frac{I_{\mathrm{mag}}}{I_{\mathrm{raw}}}\sim \frac{20\text{--}30}{1000}\approx 2\text{--}3\%.
\end{equation}
Thus, $\gtrsim 95\%$ of the raw signal in this energy range arises from non-magnetic background, while the photon contribution is at the percent level.

\noindent\textbf{Consistency between field-subtraction and polarized-neutron analysis.}
As mentioned in the introduction part of the main text, we extracted the photon signal using polarized neutrons in previous work \cite{Gao2025}. Here we validate the STHFBS protocol by comparing the resulting spectral shape with the magnetic response $M_y(E)+M_z(E)$ obtained independently from polarized-neutron spin-flip analysis reported in Ref.~\cite{Gao2025}. As shown in Fig.~\ref{fig:S_compare_pol_vs_field}, the two approaches yield consistent energy dependence in the low-energy regime: the field-subtracted spectrum captures the same characteristic dominate gapless peak/gaped shoulder and the decay into a weak higher-energy tail. The difference in the FWHM is due to the different energy resolution, and the much larger error bars in the polarized neutron results are due to the much smaller flux in the polarized experiment.

\section{Dipolar-Octupolar Pyrochlores
\label{sec:Appendix_DO_compound}}
As a result of crystal electric field (CEF) splitting in the $D_{3d}$ environment of the rare-earth ions, the low-energy manifold on each pyrochlore site of Ce$_2$Zr$_2$O$_7$ is well described by a Kramers doublet~\cite{rau2019frustrated, gingras2014quantum}. Restricting to this doublet, we introduce an effective pseudospin-$1/2$ operator $\boldsymbol{\tau}_i = (\tau_i^x,\tau_i^y,\tau_i^z)$ acting in the local frame $\{\hat{\mathbf{x}}_i,\hat{\mathbf{y}}_i,\hat{\mathbf{z}}_i\}$ (see Table~\ref{tab: Local basis}).

\begin{table}[!ht]
\caption{\label{tab: Local basis}%
Local sublattice basis vectors.
}
\begin{ruledtabular}
\begin{tabular}{ccccc}
$i$ & 0 & 1  & 2  & 3 \\
\hline
$\hat{x}_{i}$ & $\frac{1}{\sqrt{6}}\left(-2,1,1\right)$ & $\frac{-1}{\sqrt{6}}\left(2,1,1\right)$  & $\frac{1}{\sqrt{6}}\left(2,1,-1\right)$  & $\frac{1}{\sqrt{6}}\left(2,-1,1\right)$    \\
$\hat{y}_{i}$ & $\frac{1}{\sqrt{2}}\left(0,-1,1\right)$  & $\frac{1}{\sqrt{2}}\left(0,1,-1\right)$  & $\frac{-1}{\sqrt{2}}\left(0,1,1\right)$ & $\frac{1}{\sqrt{2}}\left(0,1,1\right)$  \\[2mm]
$\hat{z}_{i}$ & $\frac{1}{\sqrt{3}}\left(1,1,1\right)$ & $\frac{-1}{\sqrt{3}}\left(-1,1,1\right)$  & $\frac{-1}{\sqrt{3}}\left(1,-1,1\right)$  & $\frac{-1}{\sqrt{3}}\left(1,1,-1\right)$   \\[2mm]
\end{tabular}
\end{ruledtabular}
\end{table}

Microscopically, one first diagonalizes the CEF Hamiltonian in the atomic basis $\{\lvert J,m_J\rangle\}$ of the free-ion multiplet (for Ce$^{3+}$, $J=5/2$) to obtain a set of CEF eigenstates. The ground-state Kramers doublet $\{\lvert + \rangle, \lvert - \rangle\}$ consists of two linear combinations of these $\lvert J,m_J\rangle$ states. The effective pseudospin operators $\tau_i^\alpha$ are then defined as Pauli matrices acting within this doublet subspace:
\begin{equation}
    \tau_i^z = \tfrac{1}{2}\big(\lvert + \rangle\langle + \rvert - \lvert - \rangle\langle - \rvert\big),\quad
    \tau_i^x = \tfrac{1}{2}\big(\lvert + \rangle\langle - \rvert + \lvert - \rangle\langle + \rvert\big),
\end{equation}
and similarly for $\tau_i^y$. Any microscopic operator projected into the CEF ground-state doublet, such as components of the total angular momentum $\mathbf{J}_i$ or higher-rank multipoles, is thus represented by a $2\times 2$ matrix and can be expanded as a linear combination of the identity and the pseudospin components $\tau_i^\alpha$.

For the Ce$^{3+}$ ions in Ce$_2$Zr$_2$O$_7$, symmetry analysis shows that the three components of $\boldsymbol{\tau}_i$ do not all transform as conventional magnetic dipoles under the point-group operations. Instead, two components, $\tau_i^x$ and $\tau_i^z$, transform as magnetic dipoles, while the remaining component $\tau_i^y$ transforms as a magnetic octupole along the local trigonal axis $\hat{\mathbf{z}}_i$~\cite{huang2014quantum, gaudet2019quantum}. Consequently, the local magnetic moment projected into the ground-state doublet can be written as
\begin{equation}
    \mathbf{m}_i \simeq -\mu_B g_{zz}\,\tau_i^z\,\hat{\mathbf{z}}_i
\end{equation}
where $g_{zz}$ is the longitudinal component of the effective $g$-tensor. The octupolar components $\tau_i^{x,y}$ do not contribute linearly to $\mathbf{m}_i$.

Since the neutron couples to the magnetic dipole moment, the leading contribution to the inelastic neutron scattering cross section is governed by correlations of $\tau^z$~\cite{Gao2019, Smith2022, Smith2023, Gao2025}. In particular, the measured dynamical structure factor is proportional to the dipolar correlator $\langle \tau^z(-\mathbf{q},t)\,\tau^z(\mathbf{q},0)\rangle$.

\section{Exact Diagonalization}
\begin{figure}
    \centering
    \includegraphics[width=0.8\linewidth]{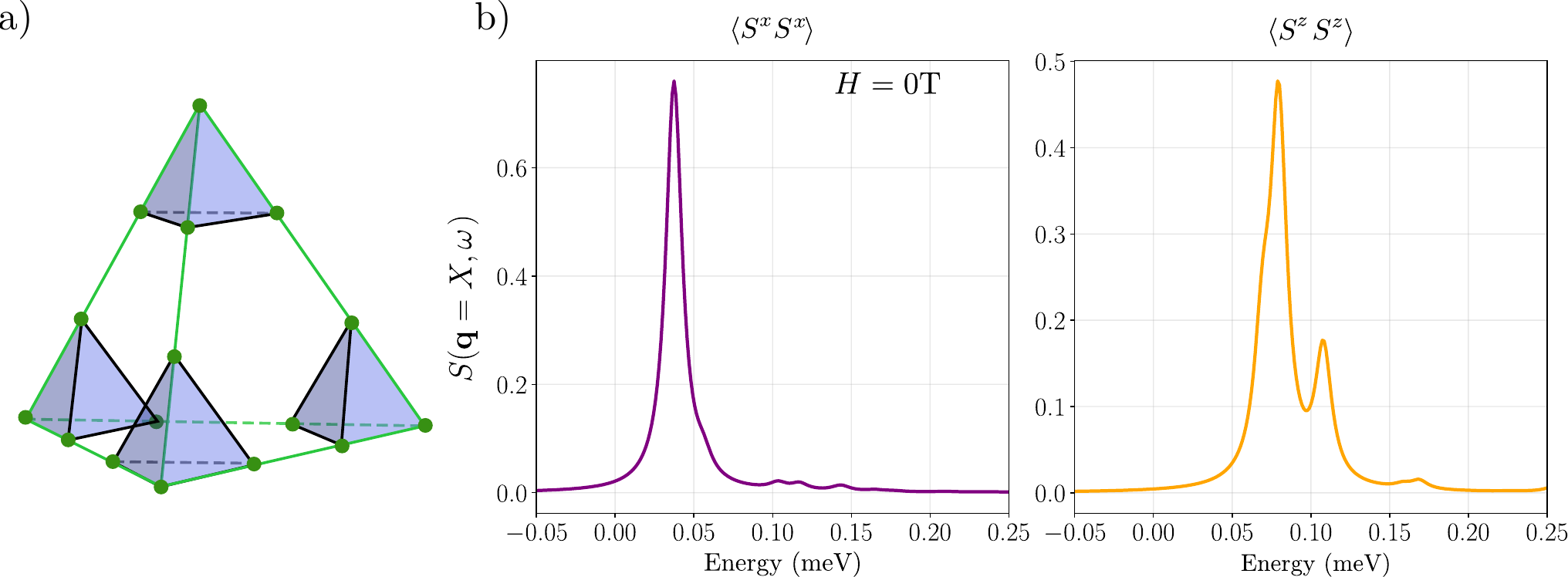}
    \caption{a) 16-site cubic conventional cluster geometry for exact diagonalization. Due to the small cluster size, shorter 4-site spin flip exchanges are possible, contributing to photon dynamics, and are highlighted in green. b) Spectral function of $\langle S^x S^x\rangle$ and $\langle S^z S^z\rangle$ at zero field with no mixing angle.}
    \label{fig:supp_ED}
\end{figure}

Exact diagonalization is performed on a 16-site periodic cubic cluster, shown in Fig.~\ref{fig:supp_ED}(a). To compute the dynamical spin structure factor (DSSF), we first obtain the ground state using the locally optimal preconditioned block conjugate gradient (LOBPCG) method~\cite{knyazev2001toward} and then evaluate the finite-frequency spin–spin correlation functions via the Lanczos continued-fraction method~\cite{gagliano1987dynamical, prelovsek2011ground}.

To assess finite-size effects, we show in Fig.~\ref{fig:supp_ED}(b) the zero-field DSSF at $\mathbf{q} = X$. On this 16-site cluster, the low-energy photonic feature appears with an intensity comparable to that of the spinon continuum, rather than being as strongly photon-dominated as expected for a $U(1)$ quantum spin ice in the thermodynamic limit. Gauge-theory analyses of QSI predict that, near pinch-point wave vectors, the low-energy spectral weight is largely controlled by the gapless emergent photon, leading to a pronounced quasielastic contribution, especially when the emergent photon velocity is small~\cite{ross2011quantum, Benton2012, Hermele2004, savary2012coulombic}. This is precisely the situation inferred for Ce$_2$Zr$_2$O$_7$, where thermodynamics and neutron scattering indicate a slow emergent photon and a sizable quasielastic signal associated with it~\cite{Gao2025}.

The apparent reduction of photon weight on the 16-site cluster should therefore be regarded as a finite-size artifact. The emergent photon is a collective mode built from coherent fluctuations of hexagonal spin-flip loops. On the minimal cubic cluster, the linear system size is only one conventional unit cell, so there is no genuine small-$|\mathbf{q}|$ regime: the smallest nonzero momenta already lie at $X$. In addition, the finite geometry admits a shorter four-site spin-flip loop that winds around the boundary of the cluster, as illustrated in Fig.~\ref{fig:supp_ED}(a) by the highlighted green lines. This loop still coherently connects two distinct two-in–two-out spin-ice configurations and thus generates an effective four-site ring exchange with amplitude $K_4 \sim \frac{J_\pm^2}{J_{zz}}$, contributing to the photon dynamics. By contrast, on the infinite pyrochlore lattice, the leading process within the spin-ice manifold is the hexagonal ring exchange with amplitude $K_6 \sim \frac{ J_\pm^3}{J_{zz}^2}$~\cite{Hermele2004, Benton2012}. As such, the artificial four-site process on the 16-site cluster is parametrically larger, $K_4 > K_6$, thereby substantially enhancing the effective photon velocity. As a result, the photon-like mode at $\mathbf{q}=X$ is pushed to much higher energy and its contribution to the quasielastic intensity is strongly suppressed compared to the thermodynamic limit.

At the same time, the cluster contains only a small number of inequivalent hexagonal loops, so the matrix elements of the extended photon mode are strongly suppressed and its intensity appears comparable to that of more local spinon excitations, in contrast to the thermodynamic-limit expectation.

In our parametrization, the photon-dominated channel is primarily carried by the $S^x$ correlations in the rotated local basis. Projecting onto the experimentally accessible dipolar component $\tau^z$ yields a neutron cross section proportional to
\begin{equation}
    \langle \tau^z \tau^z \rangle 
    = \sin^2\theta \,\langle S^x S^x \rangle 
    + \cos^2\theta \,\langle S^z S^z \rangle 
    + \sin(2\theta)\,\mathrm{Re}\,\langle S^x S^z \rangle .
\end{equation}
Because the photonic weight resides predominantly in $\langle S^x S^x \rangle$, the already underestimated photon amplitude on the 16-site cluster is further reduced in the $\tau^z\tau^z$ channel by the prefactor $\sin^2\theta$, yielding a quasielastic intensity that is significantly smaller than expected. For Ce$_2$Zr$_2$O$_7$, fits to thermodynamic and spectroscopic data suggest $\theta \simeq 0.1\pi$; with this value, $\sin^2\theta \approx 0.1$, and the photon contribution to $\langle \tau^z \tau^z \rangle$ is suppressed by nearly an order of magnitude. In order to partially compensate for this finite-size underestimation of the photon weight and obtain a more realistic quasielastic intensity in the $\tau^z\tau^z$ channel, we adopt $\theta = 0.2\pi$ in our ED calculations. This modest adjustment of $\theta$ does not change the qualitative identification of the dipolar–octupolar $U(1)$ quantum spin liquid, but it leads to a more quantitatively consistent comparison with the experimentally observed quasielastic signal.

\clearpage
\bibliography{227ref}

\end{document}